\documentclass[journal,onecolumn]{IEEEtran}
\usepackage{booktabs}
\usepackage{subfigure}
\usepackage{amsmath}                %for great looking maths
\usepackage{amsthm}                 %for great-looking theorems, definitions, etc.
\usepackage{amssymb}
\usepackage{thmtools}
\usepackage{kpfonts}
\usepackage[geometry]{ifsym}
\usepackage{dsfont}
\usepackage{multirow}
\usepackage[mathscr]{euscript}
\usepackage{graphicx}
\usepackage{color}
\usepackage[inline]{enumitem}
\usepackage{framed}
\usepackage{float}
\usepackage{longtable}
\usepackage{cite}
\usepackage{enumitem}
\usepackage{caption} % For captions under figures
\usepackage{subcaption} % For sub-figures environments
\usepackage{circuitikz}
\usepackage{tikz}

\usetikzlibrary{patterns,decorations.pathmorphing,positioning, decorations.markings}
\usepackage{xcolor}

\usepackage[colorlinks=true, citecolor=blue, linkcolor=blue]{hyperref}       %if you want hyperlinks in your pdf document!
\usepackage[font=itshape]{quoting}
\usepackage{lscape}
\usepackage{makecell}
\usepackage[section]{placeins}

\usepackage{array}
\usepackage{booktabs}
\usepackage{ragged2e} % for \RaggedRight
\usepackage{graphicx}
\usepackage{listings}
\usepackage{comment}
\usepackage{framed} % Framing content
\usepackage{nomencl} % Nomenclature package
\makenomenclature
\allowdisplaybreaks
\declaretheoremstyle[%
  headfont=\bfseries,%
  headpunct={:},%
  notefont=\normalfont\bfseries,%
  notebraces={--~}{},% punctuation before and after the note
    qed=$\blacksquare$,
]{definitionstyle}
\theoremstyle{definition}
\declaretheorem[style=definitionstyle,name=Definition]{defn}

\theoremstyle{definition}

\theoremstyle{plain}
\theoremstyle{remark}

\usepackage{nicematrix}
\NiceMatrixOptions{
code-for-first-row = \color{blue} ,
code-for-last-row = \color{blue} ,
code-for-first-col = \color{blue} ,
code-for-last-col = \color{blue}
}

\begin{document}

\title{A Conceptual Introduction to Hetero-functional Graph Theory for Systems-of-Systems}

\author{
\begin{minipage}[t]{0.3\textwidth}
\centering
\textbf{Prof. Amro M. Farid} \\
Department of Systems Engineering, \\
Charles V. Schaefer, Jr. School of Engineering and Science, \\
Stevens Institute of Technology, Hoboken, NJ, USA \\
MIT Mechanical Engineering, Cambridge, MA, USA \\
amfarid@alum.mit.edu
\end{minipage}
\hfill
\begin{minipage}[t]{0.3\textwidth}
\centering
\textbf{Amirreza Hosseini} \\
Department of Systems Engineering, \\
Charles V. Schaefer, Jr. School of Engineering and Science, \\
Stevens Institute of Technology, Hoboken, NJ, USA \\
shossein1@stevens.edu
\end{minipage}
\hfill
\begin{minipage}[t]{0.3\textwidth}
\centering
\textbf{Prof. John C. Little} \\
Charles E. Via, Jr. Department of Civil \& Environmental Engineering, \\
College of Engineering, \\Virginia Tech, Blacksburg, VA, USA \\
jcl@vt.edu
\end{minipage}
}

\maketitle

\begin{abstract}
A defining feature of twenty-first-century engineering challenges is their inherent complexity, demanding the convergence of knowledge across diverse disciplines. Establishing consistent methodological foundations for engineering systems remains a challenge—one that both systems engineering and network science have sought to address. Model-based systems engineering (MBSE) has recently emerged as a practical, interdisciplinary approach for developing complex systems from concept through implementation. In contrast, network science focuses on the quantitative analysis of networks present within engineering systems.
This paper introduces hetero-functional graph theory (HFGT) as a conceptual bridge between these two fields, serving as a tutorial for both communities. For systems engineers, HFGT preserves the heterogeneity of conceptual and ontological constructs in MBSE, including system form, function, and concept. For network scientists, it provides multiple graph-based data structures enabling matrix-based quantitative analysis. The modeling process begins with ontological foundations, defining an engineering system as an abstraction and representing it with a model. Model fidelity is assessed using four linguistic properties: soundness, completeness, lucidity, and laconicity. A meta-architecture is introduced to manage the convergence challenges between domain-specific reference architectures and case-specific instantiations. Unlike other meta-architectures, HFGT is rooted in linguistic structures, modeling resources as subjects, system processes as predicates, and operands—such as matter, energy, organisms, information, and money—as objects. These elements are integrated within a system meta-architecture expressed in the Systems Modeling Language (SysML). The paper concludes by offering guidance for further reading.

\end{abstract}

\begin{IEEEkeywords}
Model-based Systems Engineering, Network Science, Hetero-functional Graph Theory, Convergence, Ontology
\end{IEEEkeywords}

\maketitle
\section{Introduction:  Motivation for Hetero-functional Graph Theory}
One defining characteristic of twenty-first-century engineering challenges is the breadth of their scope. These societal challenges include:
\begin{enumerate}
    \item Stabilize carbon emissions
    \item Manage the nitrogen cycle
    \item Provide access to clean water
    \item Adapt to climate change
    \item Provide infrastructure for urbanized populations
    \item Feed a growing population sustainably
    \item Manage the phosphorus cycle
    \item Clean the world’s oceans of solid waste
    \item Restore and improve global biodiversity
    \item Supply human needs for energy sustainably
\end{enumerate}

At first glance, each of these aspirational engineering goals is so large and complex in its own right that it might seem entirely intractable. To adapt to each compelling challenge, we must converge knowledge that collectively spans every discipline.  And yet, these seemingly disconnected challenges are actually intertwined in Systems-of-Systems.  A convergent initiative addressing a single societal challenge is likely to be uncoordinated with, and perhaps counterproductive to, another convergent effort directed at another societal challenge.  Our single-discipline, single-problem thinking must evolve into a multi-discipline, multi-problem approach.  Fortunately, a developing consensus across several STEM (science, technology, engineering, and mathematics) fields is emerging, indicating that each of these goals is characterized by an “engineering system” that is analyzed and re-synthesized using a convergence paradigm comprising a meta-problem-solving skill set.  Instead of identifying a plethora of 21st-century societal challenges, it may be valuable to reformulate these challenges into a single convergence challenge that advances a complex system-of-systems\cite{Farid:2022:ISC-J51}.

\begin{defn}
Engineering system\cite{De-Weck:2011:00}:  A class of systems characterized by a high degree of technical complexity, social intricacy, and elaborate processes, aimed at fulfilling important functions in society.
\end{defn}

\begin{table*}[htbp]
\centering
\caption{A Classification of Engineering Systems by Function and Operand\cite{De-Weck:2011:00}}
\renewcommand{\arraystretch}{1.4}
\begin{tabular}{>{\RaggedRight}p{2.4cm} >{\RaggedRight}p{2.4cm} >{\RaggedRight}p{2.4cm} >{\RaggedRight}p{2.4cm} >{\RaggedRight}p{2.4cm} >{\RaggedRight\arraybackslash}p{2.4cm}}
\toprule
\textbf{Function/Operand} & \textbf{Living Organisms} & \textbf{Matter} & \textbf{Energy} & \textbf{Information} & \textbf{Money} \\
\midrule
\textbf{Transform} & Hospital & Blast Furnace & Engine, electric motor & Analytic engine, calculator & Bureau of Printing \& Engraving \\
\textbf{Transport} & Car, Airplane, Train & Truck, train, car, airplane & Electricity grid & Cables, radio, telephone, and internet & Banking Fedwire and SWIFT transfer systems \\
\textbf{Store}     & Farm, Apartment Complex & Warehouse & Battery, flywheel, capacitor & Magnetic tape \& disk, book & U.S. Bullion Repository (Fort Knox) \\
\textbf{Exchange}  & Cattle auction, (illegall) human trafficking & eBay trading system & Energy market & World Wide Web, Wikipedia & London Stock Exchange \\
\textbf{Control}   & U.S. Constitution \& laws & National Highway Traffic Safety Administration & Nuclear Regulatory Commission & Internet engineering task force & United States Federal Reserve \\
\bottomrule
\end{tabular}
\label{tab:functions}
\end{table*}

The challenge of convergence towards abstract and consistent methodological foundations for engineering systems is formidable.  Consider the engineering systems taxonomy presented in Table \ref{tab:functions}. It classifies engineering systems by five generic functions that fulfill human needs: \begin{enumerate*}
    \item transform,
    \item transport,
    \item store,
    \item exchange, and 
    \item control.
\end{enumerate*} On another axis, it classifies them by their operands: \begin{enumerate*}
    \item living organisms (including people),
    \item matter,
    \item energy,
    \item information, and
    \item money
    
\end{enumerate*}    This classification presents a broad array of application domains that must be consistently treated.   Furthermore, these engineering systems are at various stages of development and will likely remain so for decades, if not centuries.   And so, the study of engineering systems must equally support design synthesis, analysis, and re-synthesis while supporting innovation, be it incremental or disruptive.

Two fields in particular have attempted to traverse this convergence challenge:  systems engineering and network science.   Systems engineering, and more recently model-based systems engineering (MBSE), has evolved as a practical and interdisciplinary engineering discipline that enables the successful realization of complex systems from concept through design to full implementation \cite {SE-Handbook-Working-Group:2015:00}.  It equips the engineer with methods and tools to handle systems of ever-increasing complexity arising from greater interactions within these systems or from the expanding heterogeneity they demonstrate in their structure and function.  Despite its many accomplishments, model-based systems engineering still relies on graphical modeling languages that provide limited quantitative insight (on their own) \cite{Weilkiens:2007:00, Friedenthal:2011:00, Schoonenberg:2019:ISC-BK04}. 

In contrast, network science has developed to quantitatively analyze networks that appear in a wide variety of engineering systems.  Network science focuses on an abstract model of a system’s form, neglecting an explicit description of a system’s function\cite{Barabasi:2016:00}.  It utilizes formal graphs, comprising nodes and edges.  Nodes typically represent facilities associated with point locations, while edges represent facilities that connect the nodes.  The nodes and edges in a formal graph are described by nouns.  Because many engineering systems include multiple elements with several layers of connectivity, formal graphs are frequently generalized to create multi-layer networks.  In either case, operands are transported along edges between pairs of nodes, whether in a single layer or multiple layers.  And yet, despite its methodological developments in multi-layer networks, network science has often been unable to address the explicit heterogeneity frequently encountered in engineering systems\cite{Schoonenberg:2019:ISC-BK04,Kivela:2014:00}.   In a comprehensive review, Kivela et al. \cite{Kivela:2014:00} write: “ [The multi-layer network community] has produced an equally immense explosion of disparate terminology, and the lack of consensus (or even generally accepted) set of terminology and mathematical framework for studying is extremely problematic."

\subsection{Original Contribution}
This work provides a conceptual introduction to hetero-functional graph theory (HFGT).  On the basis of prior works, it serves as an HFGT tutorial for an MBSE and/or network science audience.  In many ways, the parallel developments of the model-based systems engineering and network science communities converge intellectually in hetero-functional graph theory (HFGT) \cite{Schoonenberg:2019:ISC-BK04}.   For the latter, it utilizes multiple graph-based data structures to support a matrix-based quantitative analysis.  For the former, HFGT inherits the heterogeneity of conceptual and ontological constructs found in model-based systems engineering, including system form, system function, and system concept.  More specifically, the explicit treatment of function and operand facilitates a structural understanding of the diversity of engineering systems found in Table \ref{tab:functions}.  In effect, HFGT quantifies many of the structural concepts found in model-based systems engineering and its associated languages (e.g., UML/SysML).  Although not named as such originally, the first works on HFGT appeared as early as 2006-2008\cite{Farid:2006:IEM-C02,Farid:2007:IEM-TP00,Farid:2008:IEM-J05,Farid:2008:IEM-J04}.  Since then, HFGT has become multiply established and demonstrated cross-domain applicability\cite{Farid:2015:ISC-J19,Schoonenberg:2019:ISC-BK04}; culminating in several recent consolidating works{\cite{Schoonenberg:2019:ISC-BK04,Farid:2022:ISC-J51, Schoonenberg:2022:ISC-J50}.

The ontological strength of hetero-functional graph theory comes from the ``systems thinking" foundations in the model-based systems engineering literature\cite{Crawley:2015:00,Schoonenberg:2019:ISC-BK04}.   In effect, and very briefly, all systems have a ``subject + verb + operand" form where the system form is the subject, the system function is the verb + operand (i.e. predicate) and the system concept is the mapping of the two to each other.   The key distinguishing feature of HFGT, relative to multi-layer networks, is its introduction of system function.  In that regard, it is more complete than multi-layer networks if the system function is accepted as part of an engineering system abstraction.  Another key distinguishing feature of HFGT is the differentiation between elements related to transformation, transportation, and storage.   In that regard, it takes great care to not \emph{overload} mathematical modeling elements and preserve lucidity.  These diverse conceptual constructs indicate multi-dimensional rather than two-dimensional relationships. Finally, HFGT has been shown to be capable of modeling an arbitrary number of engineering systems of arbitrary topology, connected to each other arbitrarily, within a system-of-systems.  

\subsection{Paper Outline}
The remainder of the paper is organized as follows.  Sec \ref{Sec:Ontology} describes the ontological underpinnings of hetero-functional graph theory.  Sec. \ref{Sec:Architectures} then discusses the use of meta-architectures as a means of overcoming the convergence challenges presented by domain-specific reference architectures and case-specific instantiated architectures.  Next, Sec. \ref{Sec:HFGT} describes some of the essential elements of HFGT.  Finally, Sec. \ref{Sec:FurtherReading} provides an entry point to further reading on HFGT.

%%\vspace{-0.1in}
\section{The Role of Systems Ontology}\label{Sec:Ontology}
HFGT draws on ontological science and can be used to compare the modeling fidelity of HFGT to the multi-layer network literature.  For example, it has been shown that HFGT overcomes eight previously identified modeling constraints in multi-layer networks \cite{Schoonenberg:2019:00, De-Domenico:2013:02}.  The modeling constraints found in multi-layer networks can be viewed from an ontological perspective.  As shown in Figure \ref{fig:Ullman's Triangle}, ontological science describes the relationship between reality (the Real Domain), the understanding of reality (the Domain Conceptualization), and the description of reality (the Language).  These general concepts are instantiated to describe the modeling of a Physical System, where the modeler’s understanding (or mental conceptualization) is the Abstraction, and the description of the abstraction is the Model.
\begin{figure}
    \centering
    \includegraphics[width=1\linewidth]{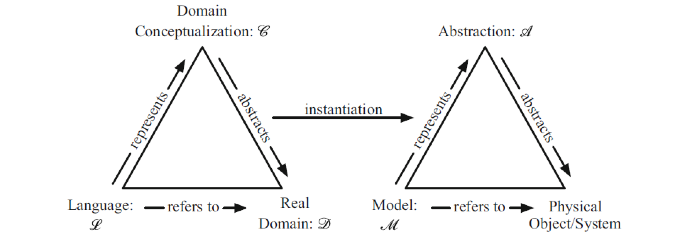}
    \caption{Ullman's Triangle\cite{guizzardi:2007:00}:  It's ontological definition.  On the left, the relationship between reality, the understanding of reality, and the description of reality.  On the right, the instantiated version of the definition.}
    \label{fig:Ullman's Triangle}
\end{figure}

The modeling process abstracts the physical system into an abstraction and represents the abstraction with a model.  The model refers to the physical system, but this reference is always indirect, as an abstraction is always made in the modeling process.  The abstraction of reality may be entirely conceptual (residing within the mind) or linguistic (residing within some predefined language).  For a model to truly represent the abstraction, the modeling primitives of the language should faithfully represent the domain conceptualization to articulate the represented abstraction.  As a result, modeling primitives directly express relevant domain concepts, creating the domain conceptualization.  The fidelity of the model with respect to the abstraction is determined by four complementary linguistic properties: soundness, completeness, lucidity, and laconicity.  When all four properties are met, the abstraction and the model have a one-to-one mapping and faithfully represent each other.

\begin{defn}[Soundness \cite{Guizzardi:2005:00}]  A language is sound with respect to a domain conceptualization if every modeling primitive in the language has an interpretation in the domain abstraction (the absence of soundness results in excess modeling primitives with respect to domain abstractions).  

\end{defn}

\begin{defn}[Completeness \cite{Guizzardi:2005:00}]  A language is complete with respect to a domain conceptualization if every concept in the domain abstraction of that domain is represented in a modeling primitive of the language (the absence of completeness results in one or more concepts in the domain abstraction not being represented by a modeling primitive).   

\end{defn}

\begin{defn}[Lucidity \cite{Guizzardi:2005:00}]  A language is lucid with respect to a domain conceptualization if every modeling primitive in the language represents at most one domain concept in the domain abstraction (the absence of lucidity results in the overload of a modeling primitive with respect to two or more domain concepts).  

\end{defn}

\begin{defn}[Laconicity \cite{Guizzardi:2005:00}]  A language is laconic with respect to a domain conceptualization if every concept in the abstraction of that domain is represented at most once in the model of that language (the absence of laconicity results in the redundancy of modeling primitives with respect to domain abstractions)

\end{defn}

It is essential to distinguish between situations where the abstraction is entirely conceptual, residing within the mind, and those where it is linguistic, residing within a predefined language.  In the case of the former, as is practiced in model-based systems engineering, the abstraction of the system is an instance of ``systems thinking” as the domain conceptualization.  Then, the (SysML) model of the system is an instance of the Systems Modeling Language (SysML) as the language.  This is the default for many engineering systems of small to moderate scale.  In the case of the latter, as is practiced in hetero-functional graph theory, the abstraction is the SysML model, and the domain conceptualization is SysML.  The hetero-functional graph model(s) are the model, and hetero-functional graph theory provides the means of representing the domain conceptualization in a mathematical language.  This is the case when the engineering system is so large and complex (e.g., a system-of-systems) that a single human mind is not able to retain the entire domain conceptualization.  Regardless of whether the abstraction is conceptual or linguistic, ontological science demands one-to-one mappings of the concepts in the mind, their graphical counterparts in SysML, and finally their mathematical counterparts in HFGT.  

\section{Instantiated, Reference, and Meta-Architectures for Systems-of-Systems}\label{Sec:Architectures}
Ullman’s triangle in Fig.\ref{fig:Ullman's Triangle} provides a basis for understanding the convergence challenge found in complex systems-of-systems.  Consider what happens when the physical/object system is now a system-of-systems.  Also, assume that the system-of-systems is a system of ``unlike” systems that have heterogeneity of type between the constituent systems.  Consequently, the real domain is a real-domain-of-real-domains that unites the constituent domains into one.  For example, the study of the food-energy-water nexus in the general (rather than a specific region) may constitute such a real-domain-of-real-domains.  It needs to be abstracted by human beings, in the mind, in a domain-conceptualization-of-domain-conceptualizations.  Similarly, that must be referred to by a language-of-languages.  Several convergence challenges immediately arise.   First, human beings are typically trained in a single-domain conceptualization, rather than multiple domains.  Indeed, it is far from clear that there even exists a single human being (let alone many) who has sufficient knowledge of the domain-conceptualization-of-domain-conceptualizations.  In the absence of such an individual, a group of individuals – each with their own domain conceptualizations – must somehow collaborate to discuss the real-domain-of-real-domains (e.g., the food-energy-water nexus independent of a specific region).  They immediately find that each domain-conceptualization comes with its associated language, and a language-of-languages emerges.  Because each of these languages was developed entirely independently to address the needs of its associated real domain, the language-of-languages is highly divergent, and a common, convergent understanding between languages is difficult to achieve.  To overcome this impasse, it is possible that the language-of-language develops a translation capability between each of the languages pertaining to each real-domain.   While this strategy is relatively straightforward for only two languages with a single translator, it does not scale when there are N real-domains that require potentially N(N-1) translators between N languages.   The remaining alternative is to invest in the development of a language-of-language that reconciles the languages together into a single common language.   

MBSE, SysML, and HFGT adopt the latter approach where a single common language serves as a language-of-languages.  The development of a single common language for a domain-conceptualization-of-domain- conceptualizations requires three types of system architectures.  As shown in Figure \ref{fig:system model}, these are the instantiated, reference, and meta-architectures.   Each of these is explained in turn.  
\begin{figure}
    \centering
    \includegraphics[width=1\linewidth]{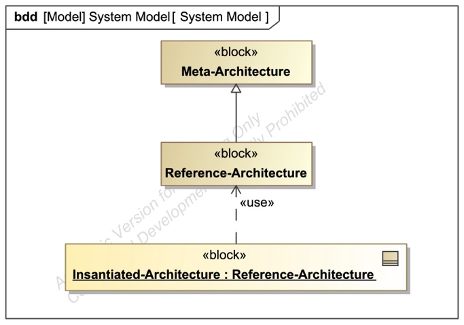}
    \caption{SysML Block Definition Diagram.  Systems architecture can be represented at three levels of abstraction: instantiated, reference, and meta-architecture.  }
    \label{fig:system model}
\end{figure}

Generally speaking, a system architecture consists of three parts: the physical architecture, the functional architecture, and the mapping of the latter onto the former in a system concept (or allocated architecture).  The physical architecture is a description of the decomposed elements of the system, without specifying the performance characteristics of the physical resources that comprise each element.  The functional architecture is a description of the system processes in a solution-neutral way, structured in serial, parallel, or hierarchical arrangements.  The system concept, as a mapping of the functional architecture onto the physical architecture, completes the system architecture.  In order to respect the Independence Axiom\cite{Suh:2001:00}, and for a hetero-functional graph model to be correctly specified, it is assumed that the entirety of any given process must be completed by a given resource.   While this assumption may seem limiting, in reality, either the process or the resource can be decomposed so that the Independence Axiom is ultimately respected.

An Instantiated Systems Architecture is a case-specific architecture that represents a real-world scenario.  At this level, the physical architecture comprises a set of instantiated resources, and the functional architecture comprises a set of instantiated system processes.  The mapping in the system concept defines which resources perform what processes.

Reference Architectures generalize instantiated system architectures.  Instead of using individual instances as elements of the physical and functional architecture, the reference architecture is expressed in terms of domain-specific classes of these instances.  In this way, the reference architecture captures the essence of existing instantiated architectures.  It also provides a vision of future needs that can guide the development of new instantiated system architectures.  Such a reference architecture facilitates a shared understanding across multiple disciplines or organizations about the current architecture and its future evolution.  A reference architecture is based on concepts that have been proven in practice.  Most often, preceding architectures are mined for these proven concepts.  The reference architecture, therefore, generalizes instantiated system architectures to define an architecture that is generally applicable in a discipline.  However, the reference architecture does not generalize beyond the domain conceptualization.

Meta-architectures further generalize reference architectures.  Instead of domain-specific elements, it is expressed in terms of domain-neutral classes.  A meta-architecture is composed of ``primitive elements” that generalize the domain-specific functional and physical elements into their domain-neutral equivalents.  While no single engineering system meta-architecture has been developed for all purposes, several modeling methodologies have been developed that span several discipline-specific domains.  In the design of dynamic systems, bond graphs \cite{Paynter:1961:00, Karnopp:1990:00, Brown:2007:00} and linear graphs \cite{Koenig:1967:00, Blackwell:1968:00, Shearer:1967:00, Kuo:1967:00, Chan:1972:00} use generalized capacitors, resistors, inductors, gyrators, and transformers as primitive elements.  In the system dynamics of business, stocks and flows are often used as primitives\cite{Forrester:1958:00,Sterman:2000:00}. Finally, formal graph theory \cite{ Steen:2010:00, Newman:2009:00} introduces nodes and edges as primitive elements.   Each of these has its respective set of applications.  However, their sufficiency must ultimately be tested by an ontological analysis of soundness, completeness, lucidity, and laconicity.  Hetero-functional graph theory, as the next section elaborates, utilizes its own meta-architecture, which, in recent years, has been shown to generalize linear graphs, bond graphs, system dynamics, and formal graph theory \cite{Ghorbanichemazkati:2024:ISC-JR03,Naderi:2024:ISC-AP97,Farid:2022:ISC-J51}.    Given the importance of ontological clarity, hetero-functional graph theory has taken special care in translating this meta-architecture from its description in SysML to its mathematical representation.  

\section{Essential Elements of Hetero-functional Graph Theory}\label{Sec:HFGT}
While a comprehensive introduction to hetero-functional graph theory can not be provided here, this section introduces the essential elements of hetero-functional graph theory in terms of its underlying meta-architecture.  Unlike other meta-architectures, hetero-functional graph theory stems from the universal structure of human language with subjects and predicates and the latter made up of verbs and objects\cite{Schoonenberg:2019:ISC-BK04,Farid:2022:ISC-J51}  An engineering system includes a set of resources $R$ as subjects, a set of system processes $P$ as predicates, and a set of operands $L$ as their constituent objects.  

\begin{defn}[System Operand \cite{SE-Handbook-Working-Group:2015:00}]
An asset or object $l_i \in L$ that is operated on or consumed during the execution of a process.  
\end{defn}

\begin{defn}[System Process \cite{Hoyle:1998:00,SE-Handbook-Working-Group:2015:00}] \label{def:CH4:process}
An activity $p \in P$ that transforms a predefined set of input operands into a predefined set of outputs. 
\end{defn}

\begin{defn}[System Resource \cite{SE-Handbook-Working-Group:2015:00}]
An asset or object $r_v \in R$ that is utilized during the execution of a process.  
\end{defn}

\begin{figure*}
    \centering
    \includegraphics[width=1\linewidth]{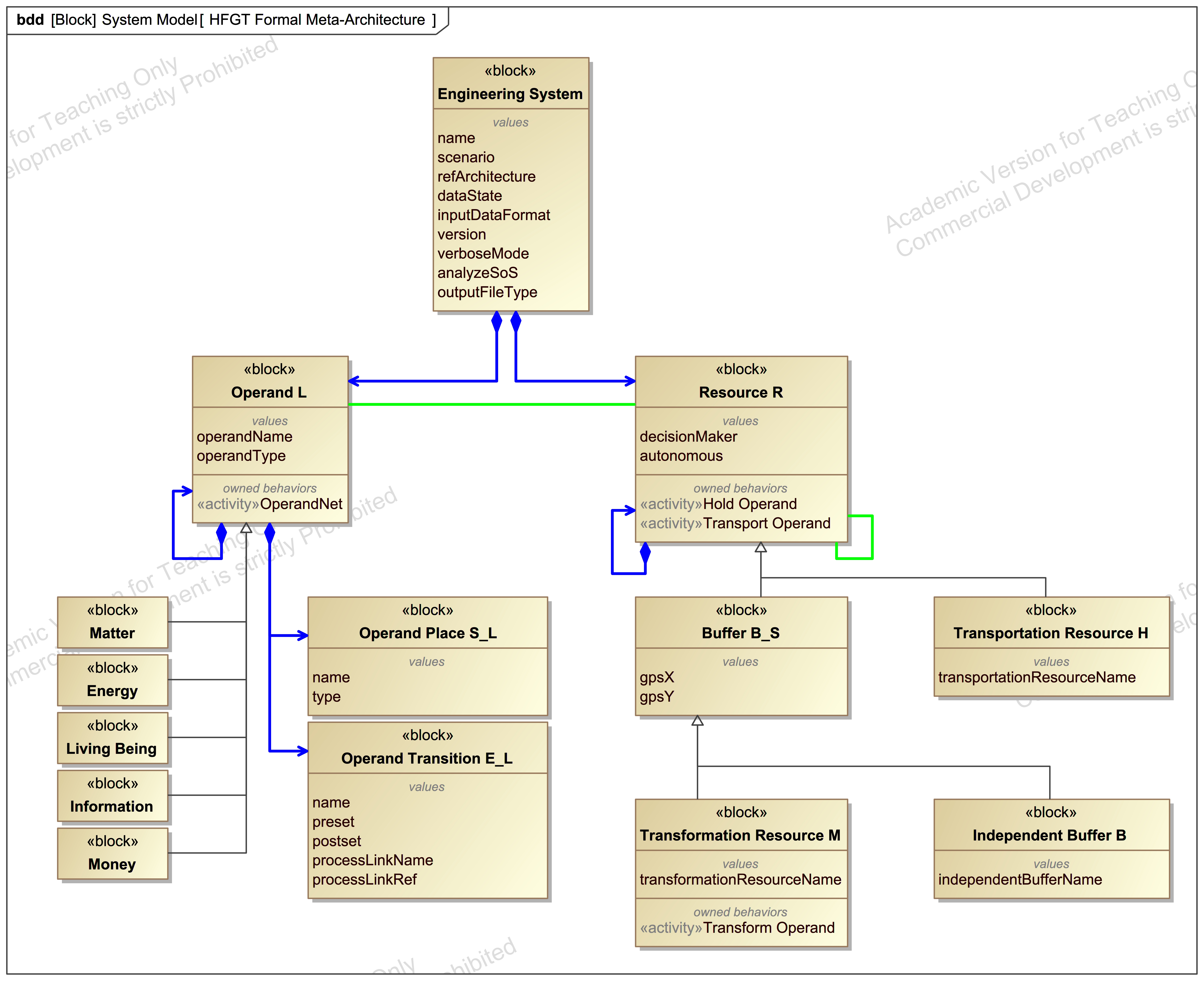}
    \caption{A SysML Block Diagram of the HFGT Formal Meta-Architecture}
    \label{fig:hfgt meta architecture}
\end{figure*}

As shown in Fig. \ref{fig:hfgt meta architecture}, these operands, processes, and resources are organized in an engineering system meta-architecture stated in the Systems Modeling Language\cite{Delligatti:2014:00,Friedenthal:2014:00,Weilkiens:2007:00}.  Importantly, operands in the engineering system have several types, including matter, energy, living beings, information, and money, so as to correspond to those operands found in Table \ref{tab:functions}.  Interestingly, it is understood that operands, in general, have some sort of state in time.  The evolution of this operand state is described by an operand net as a type of Petri net\cite{Girault:2013:00}.  

\begin{defn}[Operand Net\cite{Farid:2008:IEM-J04,Schoonenberg:2019:ISC-BK04,Khayal:2017:ISC-J35,Schoonenberg:2017:IEM-J34}]
Given operand $l_i$, an elementary Petri net $N_{l_i} = \{S_{l_i}, E_{l_i}, M_{l_i}, Q_{l_i}\}$ where:
\begin{itemize}
    \item $S_{l_i}$ is the set of places describing the operand's state.
    \item $E_{l_i}$ is the set of transitions describing the evolution of the operand's state.
    \item $M_{l_i}$ is the set of weighted arcs from places to transitions and transitions back to places, with the associated incidence matrices: 
    \[
    M_{l_i} = M^+_{l_i} - M^-_{l_i}.
    \]
\end{itemize} 
\end{defn}

The middle of Fig.\ref{fig:hfgt meta architecture} shows that the system resources $R = M \cup B \cup H$ are classified into transformation resources $M$, independent buffers $B$, and transportation resources $H$. 
\begin{defn}[Buffer \cite{Schoonenberg:2019:ISC-BK04, Farid:2022:ISC-J51}]
A resource $r_y \in R$ is a buffer $b_s \in B_s$ if and only if it is capable of storing or transforming one or more operands at a unique location in space. $B_s = M \cup B$.
\end{defn}
\noindent Additionally, the set of buffers $B_s = M \cup B$ is introduced as an important part of the theory.  HFGT inherently supports cyber-physical and socio-technical systems.  The transformation resources, independent buffers, and transportation resources can be physical in nature because they act upon matter, energy, and living beings.  Alternatively, they can be decision-making resources that act upon information and money.  Finally, Fig. \ref{fig:hfgt meta architecture} shows that resources (and their associated types) are capable of executing one or more system processes to produce a set of capabilities\cite{Schoonenberg:2019:ISC-BK04}.

\begin{figure*}
    \centering
    \includegraphics[width=1\linewidth]{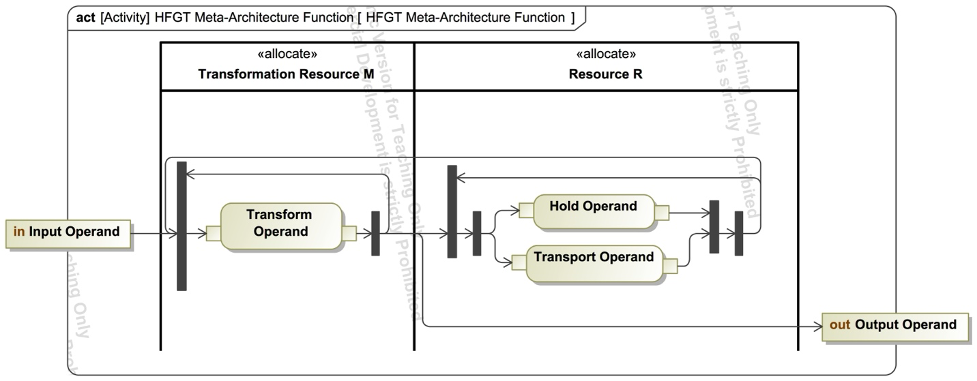}
    \caption{A SysML Activity Diagram of the HFGT Functional Meta-Architecture}
    \label{fig:hfgt meta function}
\end{figure*}

\begin{defn}[Capability \cite{Schoonenberg:2019:ISC-BK04,Farid:2022:ISC-J51,Farid:2016:ISC-BC06}] \label{def:capability}:
An action $e_{wv} \in ES$ (in the SysML sense) is defined by a system process $p_w \in P$ being executed by a resource $r_v \in R$. It constitutes a subject + verb + operand sentence of the form: \emph{Resource $r_v$ does process $p_w$}.
    
\end{defn}

The system processes $P = P_\mu \cup P_\eta$ are classified into transformation processes $P_\mu$ and refined transportation processes $P_\eta$. The latter arises from the simultaneous execution of one (unrefined) transportation process and one holding process. The relationships between the different types of system processes and the resources that can execute them is further depicted in the hetero-functional graph theory functional meta-architecture (Fig.~\ref{fig:hfgt meta function}).

Fig. \ref{fig:hfgt meta architecture} and \ref{fig:hfgt meta function} both show that only transformation resources can carry out transformation processes, while all resources can carry out refined transportation processes.  Fig. \ref{fig:hfgt meta function} also depicts the simultaneous execution of an (unrefined) transportation process and holding process in the place of a refined transportation process.  Finally, Fig. \ref{fig:hfgt meta function} shows that transformation processes and refined transportation processes can follow each other in any arbitrary sequence.  

Returning to Defn. \ref{def:capability}, it is important to recognize that while capabilities are their own distinct entities, in reality, they are formed by the allocation of a process $p_w$ to a resource $r_v$.  In Fig. \ref{fig:hfgt meta architecture}, these capabilities appear as owned behaviors of their respective blocks.  In Fig. \ref{fig:hfgt meta function}, these capabilities appear as actions in their respective swim lanes.  At an engineering system level, these allocations are described in the system concept.      

\begin{defn}[System Concept\cite{Farid:2007:IEM-TP00,Farid:2015:ISC-J19,Farid:2016:ISC-BK03}]\label{def:systemConcept}
A binary matrix $A_S$ of size $|P|\times|R|$ whose element $A_S(w,v)\in\{0,1\}$ is equal to one when action $e_{wv} \in {\cal E}_S$ (in the SysML sense) is available as a system process $p_w \in P$ being executed by a resource $r_v \in R$.
\end{defn} 
\noindent In other words, the system concept forms a bipartite graph between the set of system processes and the set of system resources\cite{Farid:2015:ISC-J19}.  

Once the engineering system’s capabilities have been defined, it is necessary to understand how they interact with one another.   These appear most clearly as the directed arrows between the actions allocated to swim lanes in Fig. \ref{fig:hfgt meta function}.  Mathematically, hetero-functional graph theory describes these functional interactions with the (third-order) hetero-functional incidence tensor ${\cal M}_\rho={\cal M}_\rho^+-{\cal M}_\rho^-$.

\begin{defn}[The Negative 3$^{rd}$ Order Hetero-functional Incidence Tensor $\widetilde{\cal M}_\rho^-$\cite{Farid:2022:ISC-J51}]
The negative hetero-functional incidence tensor $\widetilde{\cal M_\rho}^- \in \{0,1\}^{|L|\times |B_S| \times |{\cal E}_S|}$  is a third-order tensor whose element $\widetilde{\cal M}_\rho^{-}(i,y,\psi)=1$ when the system capability ${\epsilon}_\psi \in {\cal E}_S$ pulls operand $l_i \in L$ from buffer $b_{s_y} \in B_S$.
\end{defn} 

\begin{defn}[The Positive  3$^{rd}$ Order Hetero-functional Incidence Tensor $\widetilde{\cal M}_\rho^+$\cite{Farid:2022:ISC-J51}]
The positive hetero-functional incidence tensor $\widetilde{\cal M}_\rho^+ \in \{0,1\}^{|L|\times |B_S| \times |{\cal E}_S|}$  is a third-order tensor whose element $\widetilde{\cal M}_\rho^{+}(i,y,\psi)=1$ when the system capability ${\epsilon}_\psi \in {\cal E}_S$ injects operand $l_i \in L$ into buffer $b_{s_y} \in B_S$.
\end{defn} 
\noindent Interestingly, the positive and negative third-order hetero-functional incidence tensors have closed-form formulas that make their automated calculation straightforward\cite{Farid:2022:ISC-J51}.  Furthermore, these third-order incidence tensors can be straightforwardly matricized to form their associated second-order hetero-functional incidence matrices $M_\rho = M_\rho^+ - M_\rho^-$ with dimensions $|L||B_S| \times |E_S|$. This enables the formation of a hetero-functional graph and its corresponding adjacency matrix.

\begin{defn}[Hetero-functional Adjacency Matrix (Projected)]\cite{Farid:2015:ISC-J19,Farid:2015:SPG-J17,Viswanath:2013:ETS-J08,Farid:2016:ETS-J27,Schoonenberg:2017:IEM-J34}
    A square binary matrix $A_\rho$ of size $|E_S| \times |E_S|$ whose element $A_\rho(\psi_1, \psi_2) \in \{0,1\}$ is equal to one when string $z_{\{\psi_1, \psi_2\}} \in Z$ is available and exists.
\end{defn}

\begin{figure}[h]
    \centering
    \includegraphics[width=1\linewidth]{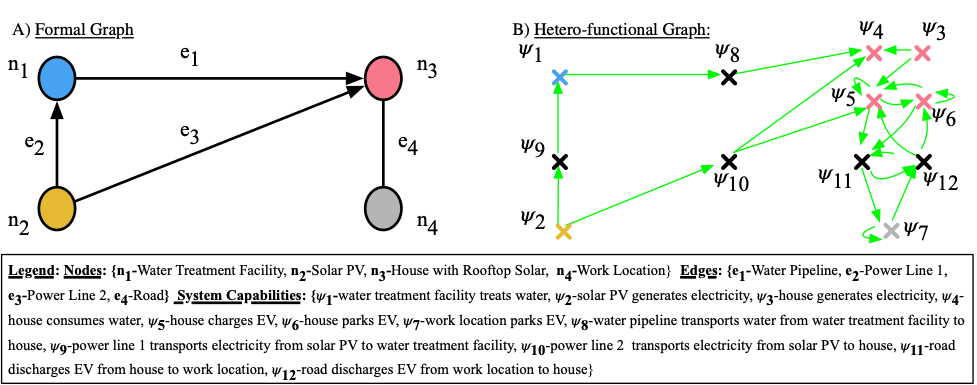}
    \caption{A Visual Comparison of a Formal Graph (FG) and a Hetero-functional Graph (HFG) Model of the Same Hypothetical System. \cite{Schoonenberg:2019:ISC-BK04}}
    \label{fig:hfgt vs formal}
\end{figure}

Interestingly, the (projected) hetero-functional adjacency matrix is straightforwardly calculated from either the second-order or third-order positive and negative hetero-functional incidence tensors\cite{Farid:2022:ISC-J51,Thompson:2022:ISC-C80}. 
\begin{align}\label{Eq:Arho1}
A_\rho(\psi_1,\psi_2) &= \sum_{i}^{|L|}\sum_{y}^{|B_S|}M_\rho^{+}(i,y,\psi_1) \cdot M_\rho^-(i,y,\psi_2)\\
A_\rho &= M_\rho^{+T} M_\rho^-
\end{align}

This highly abridged conceptual introduction to hetero-functional graph theory can also be explained graphically through the illustrative example shown in Fig. \ref{fig:hfgt vs formal}.
It illustrates the difference between a formal graph and a hetero-functional graph (HFG). The formal graph in Fig. \ref{fig:hfgt vs formal}a shows a system composed of four nodes: a water treatment facility, a solar PV panel, a house with rooftop solar, and a work location. These are connected by four edges: a water pipeline, two power lines and two roads. In contrast, Fig. \ref{fig:hfgt vs formal}b shows the associated hetero-functional graph. Instead of four nodes that represent point-like facilities, the hetero-functional graph now has 12 nodes that represent the connected system capabilities.  The water treatment facility, solar PV panel, and work location appear unchanged between the two graphs because they each have only one capability. In contrast, the house with rooftop solar provides four capabilities in the HFG. Thirdly, the edges in the formal graph are now transportation capabilities connecting nodes in the HFG. Finally, the directed edges in the HFG indicate the logical sequences of these capabilities such that if one were to follow them a “story” of capabilities would emerge. (i.e. The water treatment facility treats water ($\psi_1$) and then the water pipeline transports the water from the water treatment facility to the house ($\psi_8$)). 

Interestingly, the adjacency matrix pertaining to the formal graph can also be calculated from the third order positive and negative hetero-functional incidence tensors\cite{Farid:2022:ISC-J51,Thompson:2022:ISC-C80}.  
\begin{align}
A_{B_S}(y_1,y_2) = \bigvee_\psi^{|{\cal E}_S|}\bigvee_i^{|L|}M_\rho^{-}(i,y_1,\psi) \cdot M_\rho^+(i,y_2,\psi)
\end{align}
It has a similar mathematical form to Eq. \ref{Eq:Arho1} but collapses the dimension pertaining to the system capabilities rather than collapsing the dimension pertaining to the system buffers.  Additionally, the adjacency tensor pertaining to a multi-layer formal graph can also be calculated from the third order positive and negative hetero-functional incidence tensors\cite{Farid:2022:ISC-J51,Thompson:2022:ISC-C80}.
\begin{align}
A_{B_S}(y_1,y_2,i_1,i_2) = \bigvee_\psi^{|{\cal E}_S|}M_\rho^{-}(i_1,y_1,\psi) \cdot M_\rho^+(i_1,y_2,\psi)
\end{align}
This collapsing of dimensions in formal graphs and their multi-layer networks is the reason why both have been proven to lack ontological lucidity and completeness\cite{Farid:2022:ISC-J51}.  Hetero-functional graphs, on the other hand, have no such ontological limitation because a given capability $e_psi$ decomposes to its associated resource $r_v$ and process $p_w$ which in turn imply the associated operands $L$ and buffers $B_S$.  

\section{Further Reading on Hetero-functional Graph Theory}\label{Sec:FurtherReading}
In summary, hetero-functional graph theory (HFGT)\cite{Schoonenberg:2019:ISC-BK04,Farid:2022:ISC-J51,Farid:2016:ISC-BC06} has emerged to study many complex engineering systems and systems-of-systems.  More specifically, HFGT provides a means of algorithmically translating SysML models \cite{Delligatti:2014:00, Friedenthal:2014:00, Weilkiens:2007:00} into hetero-functional graphs and/or Petri Nets\cite{Girault:2013:00}.   Methodologically speaking, in comparison to the ontological and modeling limitations of multi-layer networks\cite{Kivela:2014:00},  HFGT has demonstrated its ability to model an arbitrary number of networked systems of arbitrary topology connected arbitrarily\cite{Schoonenberg:2019:ISC-BK04}.  Furthermore, it has demonstrated its ability to reconcile and subsequently generalize both multi-layer networks \cite{Farid:2022:ISC-J51,Thompson:2022:ISC-C80}, axiomatic design models \cite{Farid:2016:ISC-BC06, Park:2021:ISC-BC10}, system dynamics models\cite{Naderi:2024:ISC-AP97}, bond graph models \cite{Ghorbanichemazkati:2024:ISC-JR03} and linear graph models\cite{Ghorbanichemazkati:2024:ISC-JR03}. This ability to reconcile system models from disparate disciplinary sources, as well as the ability to model systems-of-systems of arbitrary topology, allows HFGT to conduct novel structural analyses \cite{Thompson:2021:SPG-J46, Thompson:2024:ISC-J55}, dynamic simulations, \cite{Khayal:2021:ISC-J48}, and optimal decision problems \cite{Schoonenberg:2022:ISC-J50}. HFGT has been applied to numerous application domains individually, including electric power\cite{ Farid:2015:SPG-J17, Thompson:2021:SPG-J46}, potable water \cite{Farid:2015:ISC-J19}, transportation\cite{Viswanath:2013:ETS-J08}, and mass-customized production systems\cite{Farid:2015:IEM-J23, Farid:2008:IEM-J05, Farid:2008:IEM-J04, Farid:2017:IEM-J13}.  Perhaps more importantly, HFGT has already demonstrated its applicability to systems-of-systems including multi-modal electrified transportation systems\cite{Farid:2016:ETS-J27, vanderWardt:2017:ETS-J33, Farid:2016:ETS-BC05}, microgrid-enabled production systems \cite{Schoonenberg:2017:IEM-J34}, personalized healthcare delivery systems\cite{ Khayal:2015:ISC-J20, Khayal:2017:ISC-J35, Khayal:2021:ISC-J48}, hydrogen-natural gas systems\cite{Schoonenberg:2022:ISC-J50}, energy-water-nexus\cite{ Farid:2024:ISC-JR04}, and the American multi-modal energy system\cite{Thompson:2024:ISC-J55}.  Interestingly, several of these systems-of-systems applications of HFGT combine both time-driven as well as discrete-event phenomena.  Additionally, HFGT has demonstrated its ability to model the supply, demand, transportation, storage, transformation, assembly, and disassembly of multiple operands in distinct locations across multiple systems over time\cite{Schoonenberg:2022:ISC-J50}.  Finally, these multiple operands include matter, energy, information, money, and living organisms\cite{Schoonenberg:2019:ISC-BK04,Farid:2022:ISC-J51} and not just energy as in engineering-centric methods like linear graphs and bond graphs.  

\section{Acknowledgments}\label{Sec:Acknowledgments}
This research is based on work supported by the Growing Convergence Research Program of the National Science Foundation under Grant Numbers OIA 2317874 and OIA 2317877.

\bibliographystyle{IEEEtran}

\bibliography{LIINESLibrary,LIINESPublications}

\end{document}